\pdfoutput=1
\documentclass[journal]{IEEEtran}
\usepackage{cite}

\ifCLASSINFOpdf
  \usepackage[pdftex]{graphicx}
  \graphicspath{{img/}}
\else
\fi
\usepackage{amsmath}
\ifCLASSOPTIONcompsoc
 \usepackage[caption=false,font=normalsize,labelfont=sf,textfont=sf]{subfig}
\else
 \usepackage[caption=false,font=footnotesize]{subfig}
\fi

\usepackage{bm}
\usepackage[binary-units=true,quotient-mode=fraction]{siunitx}
\usepackage[unicode]{hyperref}
\usepackage{cleveref}
\crefname{algorithm}{Algorithm}{Algorithms}
\Crefname{algorithm}{Algorithm}{Algorithms}
\crefname{equation}{}{}
\Crefname{equation}{Equation}{Equations}
\crefname{table}{Table}{Tables}
\Crefname{table}{Table}{Tables}
\crefname{figure}{Fig.}{Figs.}
\Crefname{figure}{Figure}{Figures}
\usepackage{url}
\usepackage{threeparttable}
\usepackage{stfloats}
\usepackage{booktabs}

\begin{document}
%
\title{Position and orientation control at micro- and mesoscales using dielectrophoresis}
%
%
%


\author{Tom\'a\v{s}~Mich\'alek and Zden\v{e}k~Hur\'ak
\thanks{T. Mich\'alek and Z. Hur\'ak were with the Faculty of Electrical Engineering, Department of Control Engineering, Czech Technical University in Prague, Karlovo namesti 13, 121 35, Prague, Czech Republic e-mail: tomas.michalek@fel.cvut.cz.}
}

%
%

\markboth{}%
{}
%



\maketitle

\begin{abstract}
The electrokinetic effect of dielectrophoresis is a promising way of inducing forces and torques on a broad class of polarizable objects at micro- and mesoscale. We introduce a non-contact micro-manipulation technique based on this phenomenon, which is capable to simultaneously position and orient a micro-object of various shapes. A visual feedback control based on a real-time optimization-based inversion of a mathematical model is employed. The presented manipulation approach is demonstrated in a series of experiments with Tetris-shaped SU-8 micro-objects performed on a chip with a quadrupolar electrode array. Using more electrodes, the method is readily extensible to simultaneous manipulation with multiple objects in biology and micro-assembly applications.
\end{abstract}


%
\IEEEpeerreviewmaketitle

\section{Introduction}
%
%
%
%

\IEEEPARstart{M}{icro-} and mesoscale manipulation is of a growing interest in various scientific and engineering disciplines. In biology, a precise and accurate control of both position and orientation is used, for example, in systems for single-cell analysis~\cite{valihrach2018platforms,gao2019recent,luo2019microfluidic,lindstrom2010overview,nilsson2009review}, single-molecule studies~\cite{zhao2013lab,liu2019dielectrophoretic}, or in micro-robotics for bioengineering applications\cite{ceylan2017mobile,yao2019microfluidic,agarwal2019dielectrophoresis}. It can also be used as a tool for controlled assembly of cell-encapsulating microgel structures for purposes of tissue engineering, where a complex organization of cells may be a vital issue.\cite{hu2012hydrogel,tasoglu2014untethered,liu2017three,cui2019multicellular} Apart from this usage, the so-called micro-assembly has many envisioned applications in the industry. The emergence of miniaturized components of the so-called {\em hybrid micro-systems} (distinguished by their superior performance and functionality) calls for effective methods of their mass production replacing the costly manual assembly.\cite{ruggeri}

A straightforward approach is to use the known solutions from present automated industrial assembly lines and to miniaturize them.\cite{diller2014three,heriban2008robotic,kim1992silicon,chronis2005electrothermally,dechev2004microassembly,volland2002electrostatically,kohl2002sma} Either manually operated or automated, all of the existing robotic micro-grippers have to cope with many challenges, the most prominent one being the adhesion effect. Since at microscale, the surface forces dominate over the volumetric ones, the manipulation principles that are known from the macro-scale, when scaled down, do not work the same way. It is generally not so problematic to grasp an object, but it is then rather hard to release it afterward. To tackle this problem, one of the approaches to the assisted release is to take an advantage of the repulsive electrokinetic forces as demonstrated by Gauthier {\em et al.}.\cite{gauthier_submerged_2006} Besides the problems with adhesion, the rigidness of the object, its material and surface properties, its specific geometry, or its fragility are the other issues that need to be considered when choosing the suitable manipulation tool.\cite{ruggeri} Furthermore, the contact-based approaches to micromanipulation are usually hard to parallelize. All of this speaks in favor of non-contact manipulation.

There exist various physical phenomena suitable for non-contact micro-manipulation, including the use of electric or magnetic fields, acoustic, hydrostatic, or optical pressure. Especially the magnetic and optical approaches~\cite{kummer_octomag_2010,haghighi_optical_2016,tung_polymer-based_2014,pawashe_two-dimensional_2012,floyd_two-dimensional_2009} are widely used. In this paper, we focus on a \emph{dielectrophoresis} (DEP). It is an electrokinetic phenomenon that fits nicely to the lab-on-chip paradigm as it enables the whole device to be miniaturized to a hand-held form and complements the referred approaches.

DEP enables us to impart both forces and torques on polarizable objects through creating and ``shaping'' the external electric field.\cite{pohl_motion_1951} This field, usually generated by a set of micro-electrodes driven by harmonic voltage signals, interacts with the charge distribution formed inside the polarized object through the well-known Coulomb forces. Their result is then the DEP force. We distinguish between several different DEP related phenomena: conventional DEP (cDEP), traveling wave DEP (twDEP), electroorientation, and electrorotation. The first two describe the force created by a gradient of the electrostatic pressure, and a gradient of the field's phase, respectively. The gradient of the phase is also responsible for the last two mentioned effects, which impose a torque acting on the object. All of these phenomena are jointly termed as \emph{general DEP} (gDEP).\cite{wang1994unified}

Feedback control of position of one or even several spherical objects using DEP, which is not limited to a finite set of cage/trap locations, has already been addressed and also experimentally demonstrated numerous times.\cite{zemanek_phase-shift_2018,zemanek_feedback_2015,kharboutly2013high,edwards2012electric} 

Just a few studies, however, deal also with orientation or even simultaneous position and orientation control of, preferably non-spherical, objects, which is necessary for micro-assembly tasks. Jiang and Mills used visual feedback to control an orientation of spherical yeast cells in a plane.\cite{jiang_planar_2015} Edwards {\em et al.} performed experiments with feedback control of orientation of gold nanowires.\cite{edwards_synchronous_2006,edwards_electric_2007} To this purpose, however, they utilize just the effect of electroorientation, which does not allow to control directly the magnitude of torque applied to the object. Our approach builds on our previous work concerning the control-oriented (fast to evaluate) mathematical model coupling both the relevant electrokinetics and hydrodynamics effects observed during electrorotation of non-spherical objects. More specifically, we presented a way for computation of the gDEP force and torque acting on an arbitrarily oriented, shaped, and heterogeneous object in fractions of a second.\cite{michalek2019control} We then extended this model by a hydrodynamic part (computable in real-time) and showed that its predictions match well the experimental observations.\cite{gdep_model_validation}

\subsection*{Contribution}
In this paper, we show how such a mathematical model can be used in feedback control of a position and an orientation of arbitrary micro-objects using gDEP. We demonstrate it in experiments with a quadrupolar electrode array (typically used for electrorotation experiments) and various Tetris-shaped micro-objects, which we steer to a randomly chosen desired locations and orientations or along predefined trajectories. Precision, accuracy, and speed of such controlled gDEP manipulation are analyzed.

Such a manipulation system has a potential to be applied for example in tissue engineering to precisely arrange microgel cell-laden structures or in drug delivery where the objects could have functionalized surfaces.

\section{Laboratory setup}
\label{sec:laboratory_setup}

\begin{figure}[!t]
\centering
  \includegraphics[width=8.3cm]{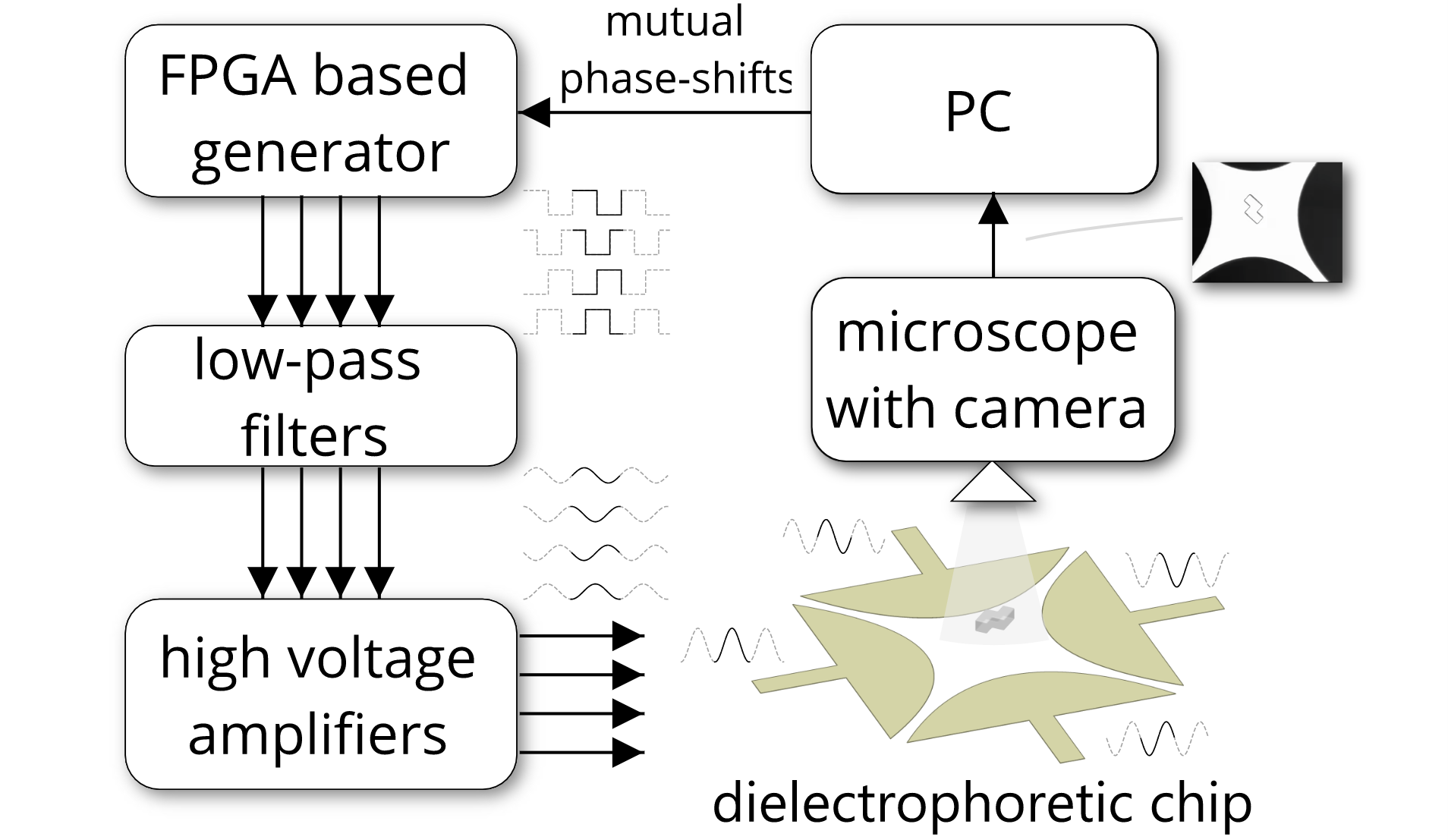}
  \caption{Diagram of the experimental setup showing the feedback loop.}
  \label{fgr:experimental_setup_diagram}
\end{figure}

The experimental setup consists of several components: the dielectrophoretic chip, microscope equipped with a camera, personal computer (PC), and a hardware for a generation of driving voltage signals (FPGA generating square waves with adjustable phase-shift, low-pass filters, and amplifiers). Their interconnection forming a feedback loop is schematically shown in~\cref{fgr:experimental_setup_diagram}. As it is indicated in the figure, the actuation happens by applying four harmonic signals differing in phase. The amplitude, as well as frequency of the voltage signals, remain fixed; all the controller can alter are their mutual phase-shifts.

The dielectrophoretic chip consists of a glass substrate with quadrupolar micro-electrodes (the arrangement showed in~\cref{fgr:experimental_setup_diagram}, which is typically used for electrorotation) fabricated on its surface. They are made by a photolithography process from gold (\SI{500}{\nano\meter}) deposited on a thin layer of chromium (\SI{20}{\nano\meter}). On the top of the chip, there is attached a plastic container holding a liquid medium with a micro-object.

\begin{figure}[!t]
\centering
  \includegraphics[width=8.3cm]{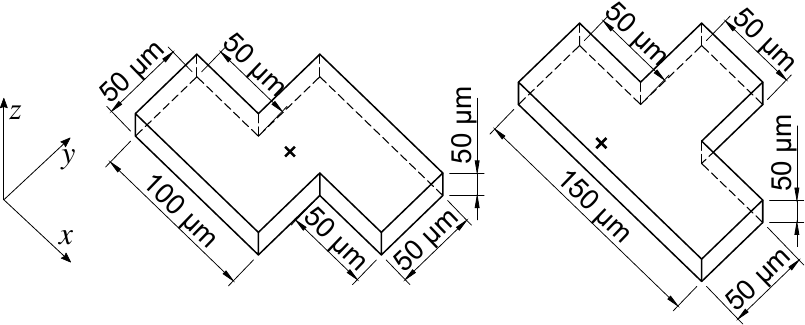}
  \caption{SU-8 micro-objects (called herein ``S/Z''-shaped, and ``T''-shaped, respectively) used in the experiments. The cross marks the point, which is used as base for measuring the object's position and orientation.}
  \label{fgr:microobjects}
\end{figure}

In this paper, we present experiments made with two different shapes of micro-objects depicted with their dimensions in~\cref{fgr:microobjects}. They are made by a photolithography process from a \SI{50}{\micro\meter} thick layer of SU-8 photoresist. All of the microfabrication was done by FEMTO-ST Institute\footnote{FEMTO-ST Institute, AS2M department Univ. Bourgogne Franche-Comté, CNRS, 24 rue Savary, F-25000 Besançon, France.}.

As the medium, we use deionized water (prepared by Water Purification System Direct-Q 3) mixed with a Polysorbate 20 (Tween 20) to reduce the surface tension of water so that we can immerse the micro-objects. The used solution has an electric conductivity of \SI{16}{\micro\siemens\per\centi\meter}. The other properties of the used materials relevant for modeling of the used physical phenomena are summarized in~\cref{tbl:material_properties}.

\begin{table}[!t]
  \renewcommand{\arraystretch}{1.3}
  \caption{\ Properties of the Used Materials}
  \label{tbl:material_properties}
  \centering
  \begin{threeparttable}
  \begin{tabular*}{0.48\textwidth}{@{\extracolsep{\fill}}lll}
    \toprule
    Property & Value & Source \\
    \midrule
    density of water soln. & \SI{998}{\kilo\gram\per\meter\cubed} & \cite{noauthor_pure_nodate} \\
    viscosity of water soln. & \SI{0.9078}{\milli\pascal\second} at \SI{25}{\celsius} & \cite{noauthor_pure_nodate} \\
    rel. permittivity of water soln. & $80$ & -- \\
    el. conductivity of water soln. & $\sim$\SI{16}{\micro\siemens\per\centi\meter} & meas.\tnote{*} \\
    density of SU-8 & \SI{1190}{\kilo\gram\per\meter\cubed} & \cite{noauthor_su-8_nodate} \\
    relative permittivity of SU-8 & $3.2$ & \cite{noauthor_microchem_nodate} \\
    electrical conductivity of SU-8 & \SI{5.556e-11}{\micro\siemens\per\centi\meter} & \cite{noauthor_microchem_nodate} \\
    \bottomrule
  \end{tabular*}
  \begin{tablenotes}\footnotesize
  \item[*] using conductivity meter Jenway 4510
  \end{tablenotes}
  \end{threeparttable}
\end{table}

The electrode array is placed under the microscope (Olympus BXFM) equipped with a long working distance objective 20$\times$/0.40 (LMPLFLN20x) and a secondary 5$\times$/0.10 (MPLN5x) objective. The first one serves as the principal one for experimental observations, while we use the second one just for calibration of coordinates (due to its greater field of view capable of also capturing the distinctive corners of the electrodes). The manipulation area is illuminated from below by a white LED panel highlighting the edges of the, otherwise transparent, micro-object. A secondary stereo microscope (Arsenal SZ 11-TH) situated right next to the primary one is used to prepare the sample before the experiment begins (see \cref{subsec:workflow} below for a description of the experimental procedure).

The video stream is captured by a digital camera (Basler acA1300-200um) and processed on a regular PC (Intel Core i5, \SI{3.30}{\giga\hertz}, 8 GB RAM, 64-bit, Win 7). An automated image processing obtains a current position and orientation of the micro-object. The PC also runs the control algorithm (described later in~\cref{sec:control} and sends the actuation commands via USB to a signal generator. Both the image processing and controller are implemented in \textsc{Matlab} software (by Mathworks) and run with a frequency of \SI{50}{\hertz}.

The phase-shifted voltage signals are generated by a custom programmed Altera DE0-Nano development board\footnote{\url{https://github.com/aa4cc/fpga-generator/}} (by Terasic Inc.). The phase resolution is \SI{1}{\degree}, and the output amplitude is \SI{3.3}{\volt}, corresponding to the logic voltage levels used by the board. We use a frequency of \SI{300}{\kilo\hertz}, which is higher than the ROT peak of \SI{3.74}{\kilo\hertz} for the given combination of medium/object materials. Still, the achievable strength of gDEP forces and torques is sufficient, and yet we avoid the unwanted effects of low-frequency electroosmosis. The generated square waves are then filtered through a pair of RLC filters in series ($R$=\SI{500}{\ohm}, $L$=\SI{100}{\micro\henry}, $C$=\SI{1.5}{\nano\farad}) with a cut-off frequency around \SI{411}{\kilo\hertz} to remove the higher-order harmonics, whose influence is not modeled for the reasons of simplicity. Another capacitor in series (\SI{100}{\nano\farad}) is used to remove the DC offset. The filtered signals are then connected to the inputs of four custom-made high-speed power amplifier modules QA210 (by Quintenz Hybridtechnik). Since the gains of these modules have a fixed value of $50$, we adjust the amplitudes of the inputs by simple voltage dividers. The output amplitudes are this way set to be around \SI{38}{\volt}.

\section{Control algorithm}
\label{sec:control}

In order to automatically manipulate the objects in the desired way, we need to measure their current position and orientation continuously. As described above, this is done by grabbing the image by the camera mounted on a microscope and sending it to a PC running the image processing and control algorithm. Based on the difference between the actual and the desired position and orientation of the object (decided, for example, by a human operator), a desired object's translational and rotational velocity vectors are computed.

Using the hydrodynamic model, we then compute the corresponding drag forces and torques that we need to overcome, taking into account the specific orientation of the object. Taking their negative and subtracting the sedimentation force arising due to gravity and buoyancy gives us the force and torque we need to exert by gDEP. The hardest (and most computationally intensive) part of the problem is to compute the appropriate phase-shifts of the voltage signals that would accomplish this. Since the model described in our previous work\cite{michalek2019control} gives force and torque based on voltages, we need to perform the model inversion. Only then we will finally obtain the parameters of voltage signals that are afterward applied to the electrodes. This whole fully automatic process repeats with a frequency of~\SI{50}{\hertz}. We will describe the mentioned subproblems in the following subsections in greater detail.

\subsection{Computer vision}

The actual position and orientation of the object are in real time automatically extracted from the image frames acquired by the camera on the microscope.

\begin{figure}[!t]
\centering
  \includegraphics[width=8.3cm]{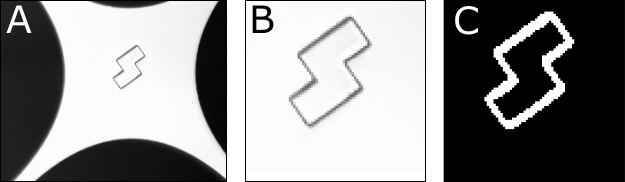}
  \caption{(A) Down-sampled image of the whole manipulation area obtained from a camera on a microscope. (B) Cropped segment of the image containing the object of interest. (C) Binary image showing the boundary of the object.}
  \label{fgr:computer_vision}
\end{figure}

The captured image is at first down-sampled to $256\times205$ and subsequently cropped to a size of $80\times80$ pixels containing just the object of interest, which is shown in~\cref{fgr:computer_vision}(A),(B), respectively. The cropping window is centered at the location where the object was detected last time. Although it still slightly lags behind the actual position of the object, its size is chosen so that the object never leaves it. This considerable reduction of the image size makes its subsequent processing much faster.

The local variance threshold-based edge detector is then used to create a binary image (shown in~\cref{fgr:computer_vision}(C)) discriminating, which part of the scene is the object's boundary and which it is not. The largest continuous blob of pixels making the object's boundary is used for further processing, making it robust against the occasional presence of small dust particles. Using the thresholded image also helps the detection algorithm to better cope with possible gradients in scene illumination.

\begin{figure}[!t]
\centering
  \includegraphics[width=8.3cm]{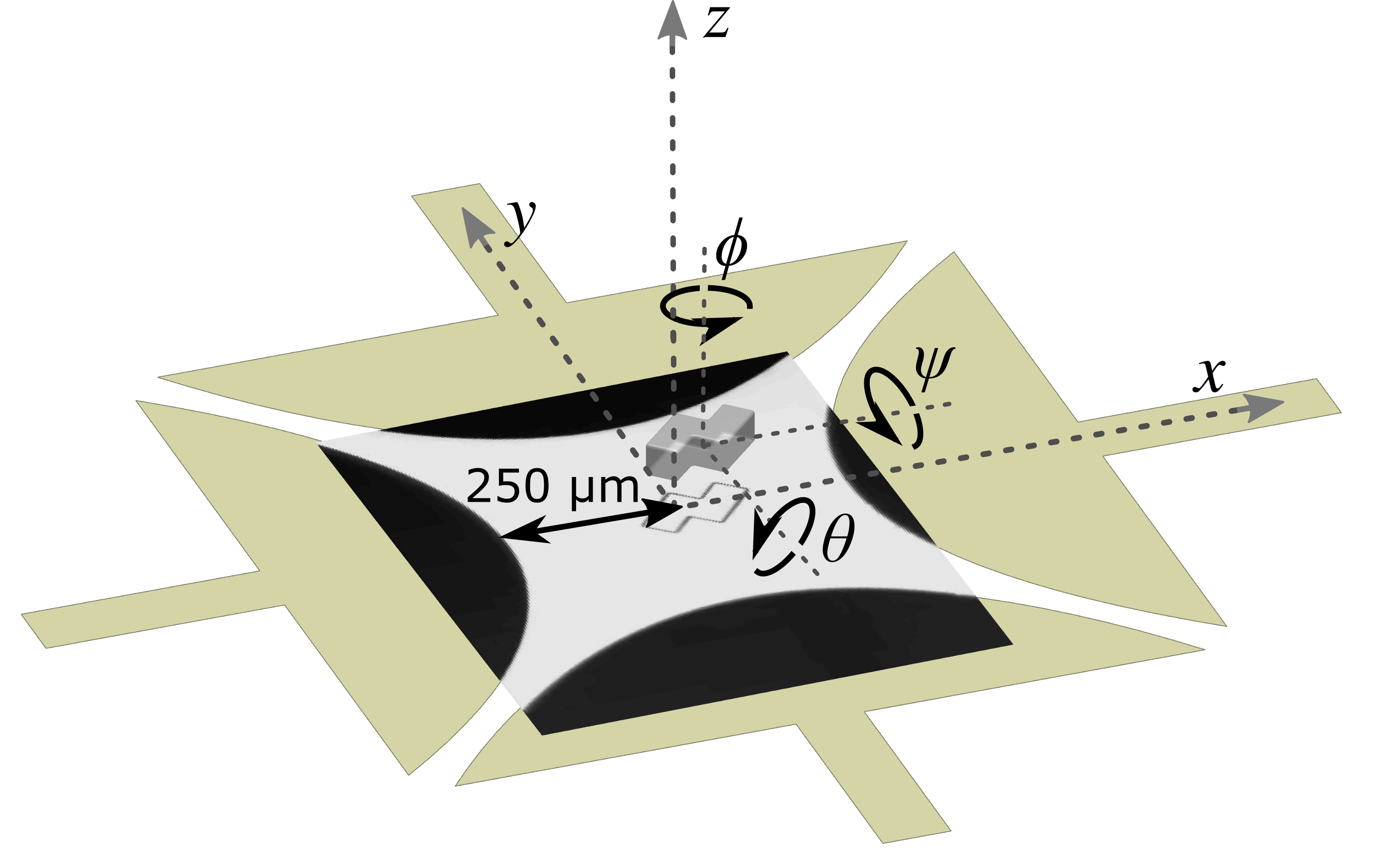}
  \caption{Definition of the used system of coordinates and directions of rotations.}
  \label{fgr:system_of_coordinates}
\end{figure}

We then use the second-order central moments of the binary image to compute the position and orientation of the object and express them in a coordinate system defined in~\cref{fgr:system_of_coordinates}.\cite{jahne2004practical} The average time of processing one frame is \SI{1.2}{\milli\second}.

\subsection{Determination of the needed force and torque}
As stated above, the first step is to determine the desired vectors of translational and rotational velocities, $\bm{v}$ and $\bm{\omega}$, respectively. We use a proportional regulator, which makes them simply proportional to the actual errors in achieved position and orientation, respectively:
\begin{equation}
\label{eq:proportional_regulator}
\begin{split}
\bm{v} &= k_v\left(\bm{r}_\mathrm{ref}-\bm{r}\right)\mathrm{,} \\
\bm{\omega} &= k_{\omega}\left(\bm{\phi}_\mathrm{ref}-\bm{\phi}\right)\mathrm{,}
\end{split}
\end{equation}
where $\bm{r}$ and $\bm{\phi}$ are the actual measured position and orientation, respectively. Analogously, $\bm{r}_\mathrm{ref}$ and $\bm{\phi}_\mathrm{ref}$ are the reference position and orientation, respectively. The scalar gains were chosen to be $k_v=50$ and $k_{\omega}=10$, respectively, leading to neither too mild nor too aggressive behavior. Based on (5) from \cite{gdep_model_validation} (originally derived in~\cite{happel_low_1965}) and the discussion at the beginning of~\cref{sec:control}, the gDEP forces and torques that should enforce such a motion are
\begin{equation}
\label{eq:hydrodynamic_model}
\begin{split}
{\bm F}_\mathrm{ref} &= \mu\mathrm{\bf K}{\bm v} + \mu\mathrm{\bf C}_O^{\mathrm T}\bm\omega-{\bm F}_{\mathrm{sed}} \mathrm{,} \\
{\bm T}_\mathrm{ref} &= \mu\mathrm{\bf C}_O{\bm v} + \mu{\bm \Omega}_O\bm\omega \mathrm{,}
\end{split}
\end{equation}
where $\mu$ is the dynamic viscosity of the liquid medium, $\mathrm{\bf K}$, $\mathrm{\bf C}_O$, and ${\bm \Omega}_O$ are {\em translational}, {\em coupling} and {\em rotational} tensors, respectively of the specific micro-object as described in \cite{gdep_model_validation}. ${\bm F}_{\mathrm{sed}}=\left[ 0, 0, \left( \rho_{\mathrm{m}}-\rho_{\mathrm{o}} \right)Vg \right]^\mathrm{T}$ is the sedimentation force incorporating both the buoyancy and gravity effect with $\rho_\mathrm{o}$ and $\rho_\mathrm{m}$ representing the density of the object and of the liquid medium respectively, $V$ being the object's volume, and $g$ denoting the gravitational acceleration.

\subsection{Model inversion using optimization}

The previously developed control-oriented model\cite{michalek2019control} provides us with a way to compute the gDEP force and torque acting on an object at a specified location and under the effect of particular voltages in fractions of a second. As it was already mentioned above, for control purposes, we need to invert the model so that it gives us the parameters of voltage signals provided the desired force and torque as the inputs.

Unfortunately, there does not exist any simple analytical inversion of the model, and we have to formulate it as a numerical optimization problem. For solving it, we need to repeatedly evaluate the force and torque at a given position many times (in our implementation up to $\sim2500$) for various input voltage signals. Since this has to be done in every control period, which is itself merely $T=\SI{20}{\milli\second}$, the model formulation from~\cite{michalek2019control} can not be directly used for this purpose.

We can, however, utilize the principle of superposition holding for electric potential (and its spatial derivatives) and reformulate the model in the same way as we showed in~\cite{zemanek_phase-shift_2018}, this time including not only the expression for force but also torque and using higher-order multipolar moments (up to the 5th order). We evaluate all of the spatial derivatives of the electric field as well as all of the multipolar moments appearing in (4-5) from~\cite{michalek2019control} for a set of scenarios. In each of them, an electric potential of \SI{1}{\volt} is applied to one of the electrodes (every time a different one), while the rest of them is kept grounded. Using this basis of solutions and expressing (4-5) from~\cite{michalek2019control} component-wise in a computer science convention, where a vector is considered to be an $n$-tuple of numbers, we get for every spatial component of force and torque a simple quadratic form
\begin{align}
F_\mathrm{a}&={\tilde{\bm u}}^\mathrm{T}\mathrm{\bf P}_\mathrm{a}\left(x,y,z,\phi,\theta,\psi\right){\tilde{\bm u}}\mathrm{,}~~~\mathrm{a}\in\left\{x,y,z\right\}\mathrm{,} \\
T_\mathrm{a}&={\tilde{\bm u}}^\mathrm{T}\mathrm{\bf Q}_\mathrm{a}\left(x,y,z,\phi,\theta,\psi\right){\tilde{\bm u}}\mathrm{,}~~~\mathrm{a}\in\left\{x,y,z\right\}\mathrm{.}
\end{align}
Here, ${\bm{\tilde u}}$ is a vector of phasors representing the harmonic voltage signals applied to the individual electrodes and $\mathrm{\bf P}$ and $\mathrm{\bf Q}$ are the position and orientation dependent matrices. Such model formulation enables us to calculate force and torque at a given position for many possible variations of voltage parameters in almost no time.

Some $\SI{1.2}{\milli\second}$ are needed to compute matrices $\mathrm{\bf P}$ and $\mathrm{\bf Q}$ for our case of four electrodes and multipoles up to the 5th order. Generally, the time $t$, which is needed, depends linearly on the number of electrodes $n\in\mathcal{Z}$ ($t=0.29n+\SI{0.083}{\milli\second}$), linear regression made from measurements on the PC used for control). 

The biggest computational bottleneck is expressing the source electric field (and its spatial derivatives) in a rotated coordinate frame aligned with the micro-object of interest, which is needed for computation of multipolar moments (for details explaining why this operation is necessary, see~\cite{michalek2019control}). In our implementation, we used a chain rule to derive the specific analytical expressions (and simplified them using a \texttt{matlabFunction}), telling us how to make this transformation. Using just-in-time (JIT) compilation in Matlab, we achieved even higher speeds than with a compiled MEX version of the same. In other cases (outside of MATLAB environment), it may be better to treat the spatial derivatives of the electric field as tensors of ascending order and utilize algorithms for (rotational) transformations of a tensor.

Having the gDEP model in a such (computationally) convenient form, we can formulate its inversion as an optimization problem. Since not all of the forces and torques are feasible to achieve, we came up with the following optimization task formulation trying to minimize various weighted error representations:
\begin{equation} \label{eq:optimization_problem}
    \begin{aligned}
    & \underset{{\bm {\tilde u}}=\left[{\tilde u}_1,\ldots,{\tilde u}_4 \right]^\mathrm{T}}{\text{minimize}}
    & & \frac{\bm{w}^\mathrm{T}\bm{e}}{\|\bm{w}\|}, \\
    & \text{subject to}
    & & \left| {\tilde u}_i \right| = U, \\
    & & & \angle {\tilde u}_i \in \left\{ 0, \frac{2\pi}{360}, \ldots, 359\frac{2\pi}{360} \right\}, \\ 
    & & & i = 1, \ldots, 4,
    \end{aligned}
\end{equation}
where $\bm{w}=\left[10,1,10,1\right]^\mathrm{T}$ is the vector of weights and $\bm{e}$ is the vector of considered errors whose individual elements are as follows: $e_\mathrm{1}$ is the percentage error in the torque direction computed as $e_\mathrm{1} = 100/\pi\cdot\arccos{\frac{{\bm T}^\mathrm{T}{\bm T}_\mathrm{ref}}{{\|{\bm T}\|\|{\bm T}_\mathrm{ref}\|}}}$, $e_\mathrm{2}$ is the percentage error in the torque magnitude computed as $e_\mathrm{2} = 100\frac{{\bm T}-{\bm T}_\mathrm{ref}}{\|{\bm T}_\mathrm{ref}\|}$, $e_\mathrm{3}$ is the percentage error in the direction of the force projected to $xy$-plane computed as $e_\mathrm{3} = 100/\pi\cdot\arccos{\frac{F_\mathrm{x}F_\mathrm{ref,x}+F_\mathrm{y}F_\mathrm{ref,y}}{\sqrt{(F_\mathrm{x}^2+F_\mathrm{y}^2)(F_\mathrm{ref,x}^2+F_\mathrm{ref,y}^2)}}}$, and finally $e_\mathrm{4}$ is the percentage error in the $z$-component of the force computed as $e_\mathrm{4} = 100\frac{F_\mathrm{z}-F_\mathrm{ref,z}}{\|F_\mathrm{ref,z}\|}$. Finally, $U=38$ represents the fixed amplitude of the voltage signals.

To solve this optimization task, we used a {\em simulated annealing} algorithm implemented according to \cite{kirkpatrick1983optimization}. It can naturally deal well with the discreteness of the set of plausible inputs, and it was much faster then the other tested global optimization solvers or their combinations (pattern search, Nelder-Mead, and BFGS Quasi-Newton method with a cubic line search procedure). The average solution time of \cref{eq:optimization_problem} is \SI{2.7}{\milli\second}.

\subsection{Achievable forces and torques}

\begin{figure}[!t]
\centering
  \includegraphics[width=8.3cm]{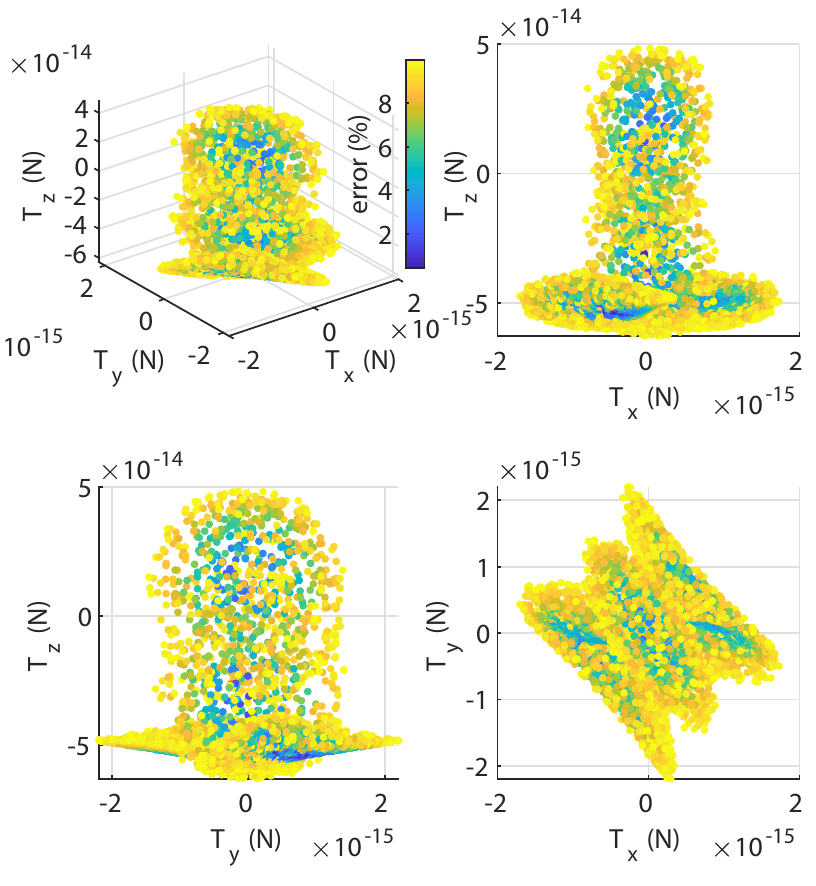}
  \caption{Achievable (feasible) torques for object located at $x=\SI{0}{\micro\meter}$, $y=\SI{0}{\micro\meter}$, $z=\SI{100}{\micro\meter}$ with a gDEP force counteracting the sedimentation force.}
  \label{fgr:feasible_torques}
\end{figure}

Not all of the forces, torques, and especially their specific combinations that we are demanding are feasible. \Cref{fgr:feasible_torques} show an estimate of the achievable torques rendered for one particular case of ``S/Z''-shaped object located at $x=y=\SI{0}{\micro\meter}$, and $z=\SI{100}{\micro\meter}$ oriented with $\phi=\theta=\psi=\SI{0}{\radian}$, while the gDEP force counteracts the sedimentation force. The color represents the percentage difference from this desired force. It is apparent that the mere four electrodes of the quadrupolar electrode array limit freedom of motion control severely. Still, we can achieve a lot as will be demonstrated experimentally, later on, in~\cref{sec:experimental_results}. Different electrode arrays with more electrodes should achieve even better results in the future.

\section{Experimental results}
\label{sec:experimental_results}
In this section, we describe the used experimental procedure and present the measured data from tasks of the simultaneous position and orientation control, and a trajectory following.

\subsection{Workflow of experiments}
\label{subsec:workflow}

Before the beginning of every experiment, we measured the temperature of the prepared liquid medium, based on which the viscosity parameter $\mu$ of the model is adjusted.

We take advantage of the long working distance, three-dimensional view, and the zoom lenses to manually position the object in between the electrodes (using a sharpened tip of a pipette), and fill up the container with the medium. We then cover the experimental chamber by a cover glass, so that the surface of the liquid medium is flat and to prevent any further contamination of the experimental chamber by dust from the surrounding.

The electrode array is then carefully transferred below the primary microscope. Since it is not possible to repeatedly achieve exactly the same position (with a micrometer precision) of the array w.r.t. the field-view of the microscope, it is always necessary to calibrate the coordinate system. A semiautomated procedure guides the user to select a few significant and well distinguishable points in the camera image (corners of the electrode array) for computation of the image transformation. 
For this purpose, we use the $5\times$ magnification since the $20\times$ objective captures mainly just the space between the electrodes where nothing other, but the object is located. For this higher magnification, the obtained transformation has to be therefore adjusted afterward by scaling and by adding some small empirically determined offset in position.

An experiment starts with a short period of pure electrorotation ($90^\circ$ mutually phase-shifted harmonic signals are applied to the electrodes) so that the object lifts off from the ground, centers itself in the space between the electrodes, and achieves its steady levitation height.

Since we are observing the scene from the top, the current experimental arrangement does not allow us to measure the levitation height continuously and use it for control purposes. We can, however, at least get its estimate at the beginning of an experiment and then require our controller to maintain the levitation height constant (at least in an open-loop regime). The same technique was already successfully deployed in~\cite{zemanek_phase-shift_2018}. Using the technique of axial distance measurement~\cite{visser1992refractive} developed in the field of 3D microscopy, we estimated the initial $z$-coordinate of the object to be approximately \SI{100}{\micro\meter}.

In the following subsections, we will present some of the obtained experimental data demonstrating the capabilities of the manipulator.

\subsection{Position and orientation control}
\begin{figure}[!t]
\centering
  \includegraphics[width=8.3cm]{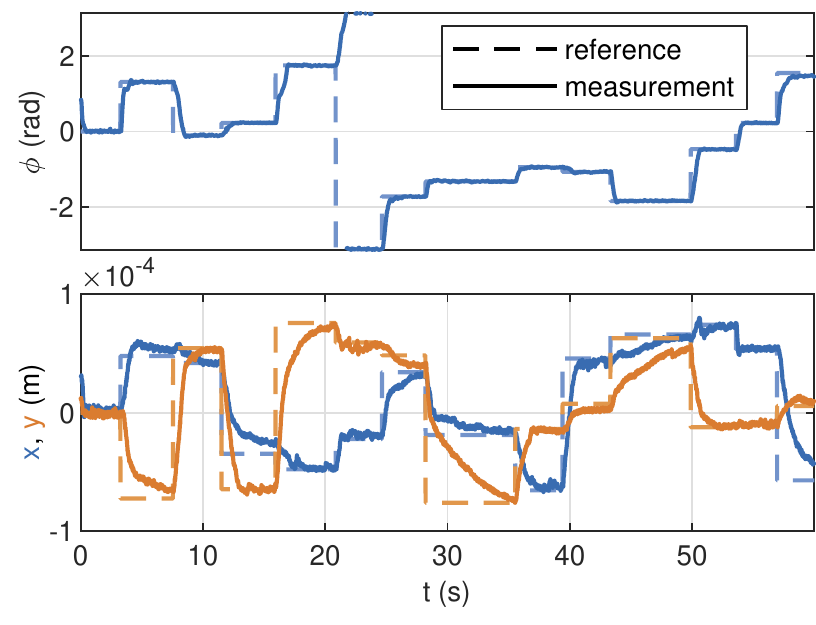}
  \caption{Results of the experiment with the ``S/Z''-shaped micro-object moving between several randomly chosen locations and orientations.}
  \label{fgr:experiment_sz}
\end{figure}

\begin{figure}[!t]
\centering
  \includegraphics[width=7cm]{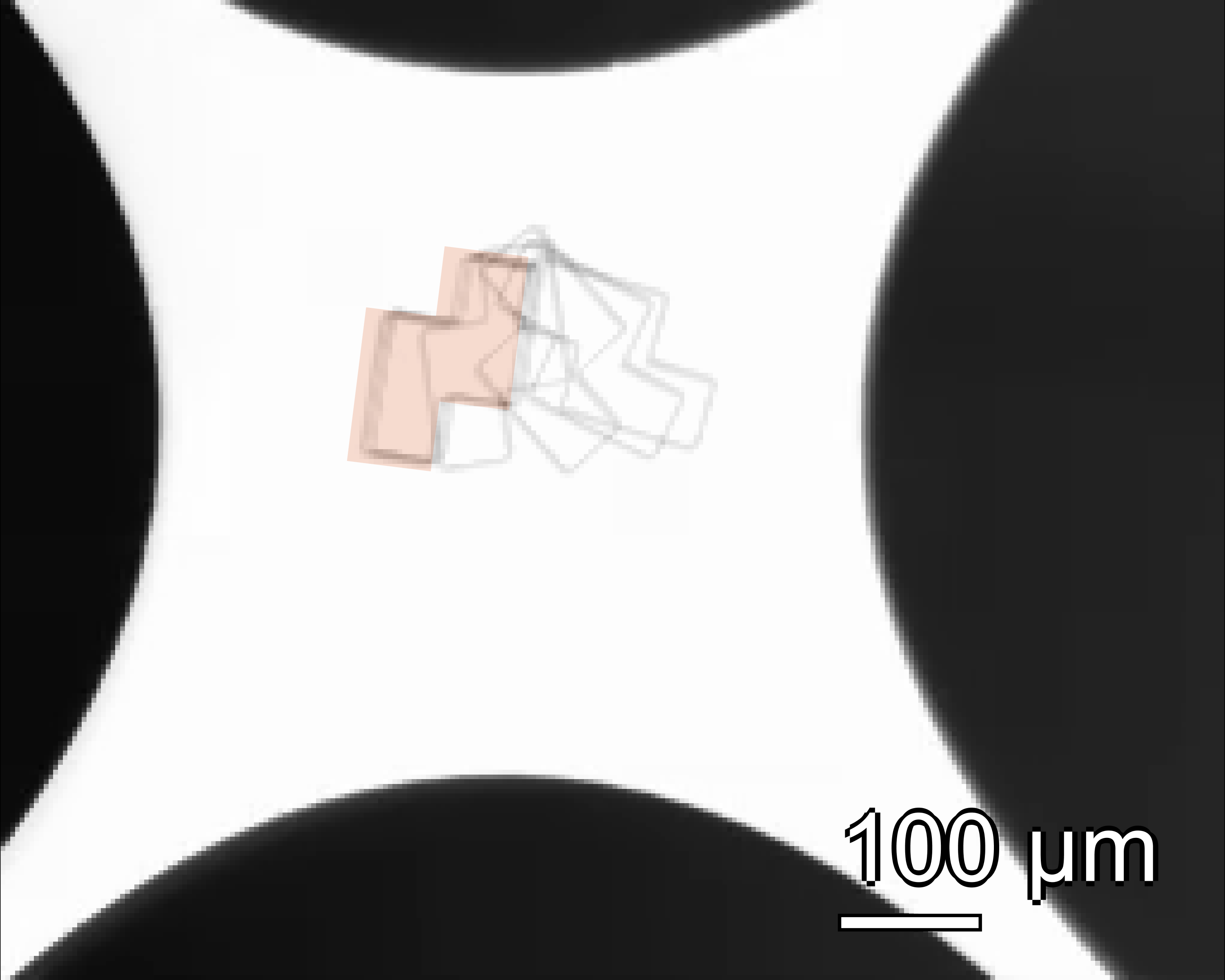}
  \caption{Fusion of several camera frames indicating the object's motion (with a time-step of \SI{250}{\milli\second}). The red silhouette represents the desired goal position and orientation of the object.}
  \label{fgr:phased_motion_sz}
\end{figure}

\begin{figure}[!t]
\centering
  \includegraphics[width=8.3cm]{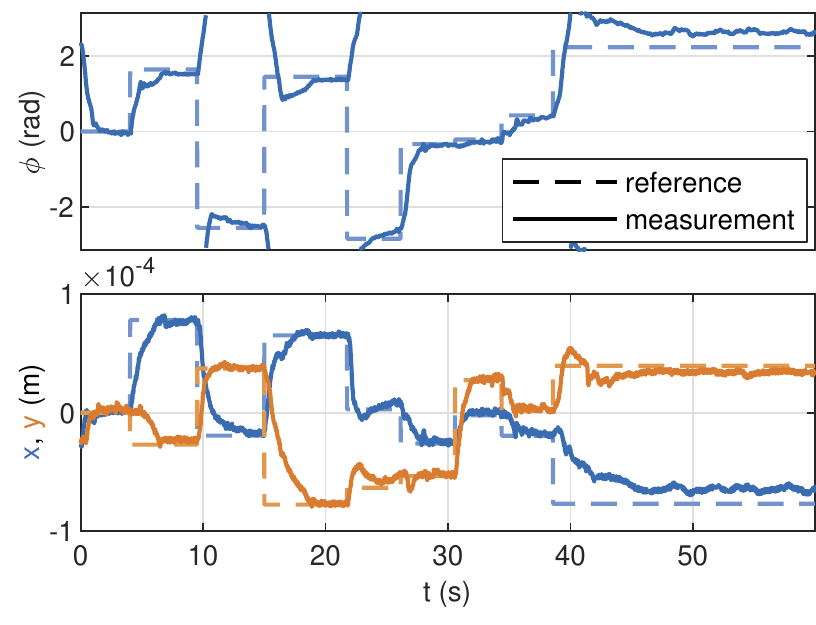}
  \caption{Results of the experiment with the ``T''-shaped micro-object moving between several randomly chosen locations and orientations.}
  \label{fgr:experiment_t}
\end{figure}

In the experiment presented in~\cref{fgr:experiment_sz}, the ``S/Z''-shaped object was steered in between several randomly chosen reference positions and orientations. These were always changed \SI{3}{\second} after the object achieved less then a \SI{20}{\micro\meter} and $\frac{\pi}{16}\,\mathrm{rad}$ error in its position and orientation, respectively. \Cref{fgr:phased_motion_sz} shows a phased-motion created by a fusion of several camera frames being all \SI{250}{\milli\second} apart from each other representing a transition of the object between two reference positions (from approximately \SI{7.54}{\second} to \SI{9.26}{\second} of experimental footage). \Cref{fgr:experiment_t}~show the same experimental procedure for the ``T''-shaped object. From the referred graphs, its is obvious that the farther the goal position of the object is from the center of the manipulation area, the harder (and thus also slower) it is for the manipulator to achieve it.

\subsection{Trajectory following}
\begin{figure}[!t]
\centering
  \includegraphics[width=8.3cm]{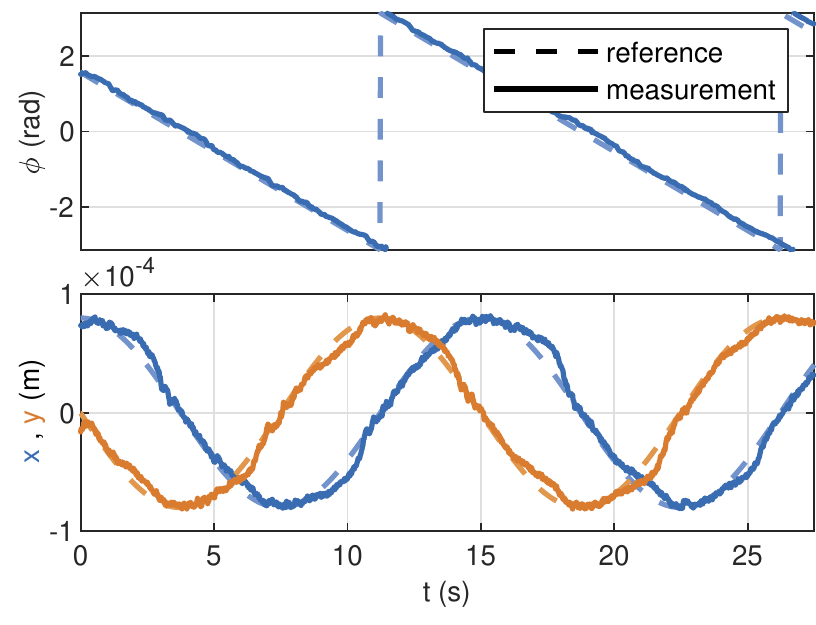}
  \caption{Results of the experiment with the ``S/Z''-shaped micro-object following a prescribed circular trajectory.}
  \label{fgr:experiment_circ_traj}
\end{figure}

By defining a sequence of mutually close reference positions, we can force an object to follow some prescribed trajectory by requiring it to pass through all points consecutively. In the experiment presented in~\cref{fgr:experiment_circ_traj}, the ``S/Z''-shaped object was steered around a circle while orienting itself so that it always remained tangent to the prescribed path.

A video from all of the above mentioned experiments is also available on-line \footnote{\url{https://youtu.be/SBepX_Xk1BM}}.

\subsection{Precision, accuracy and speed}
\begin{figure*}[!t]
\centering
  \includegraphics[width=17.1cm]{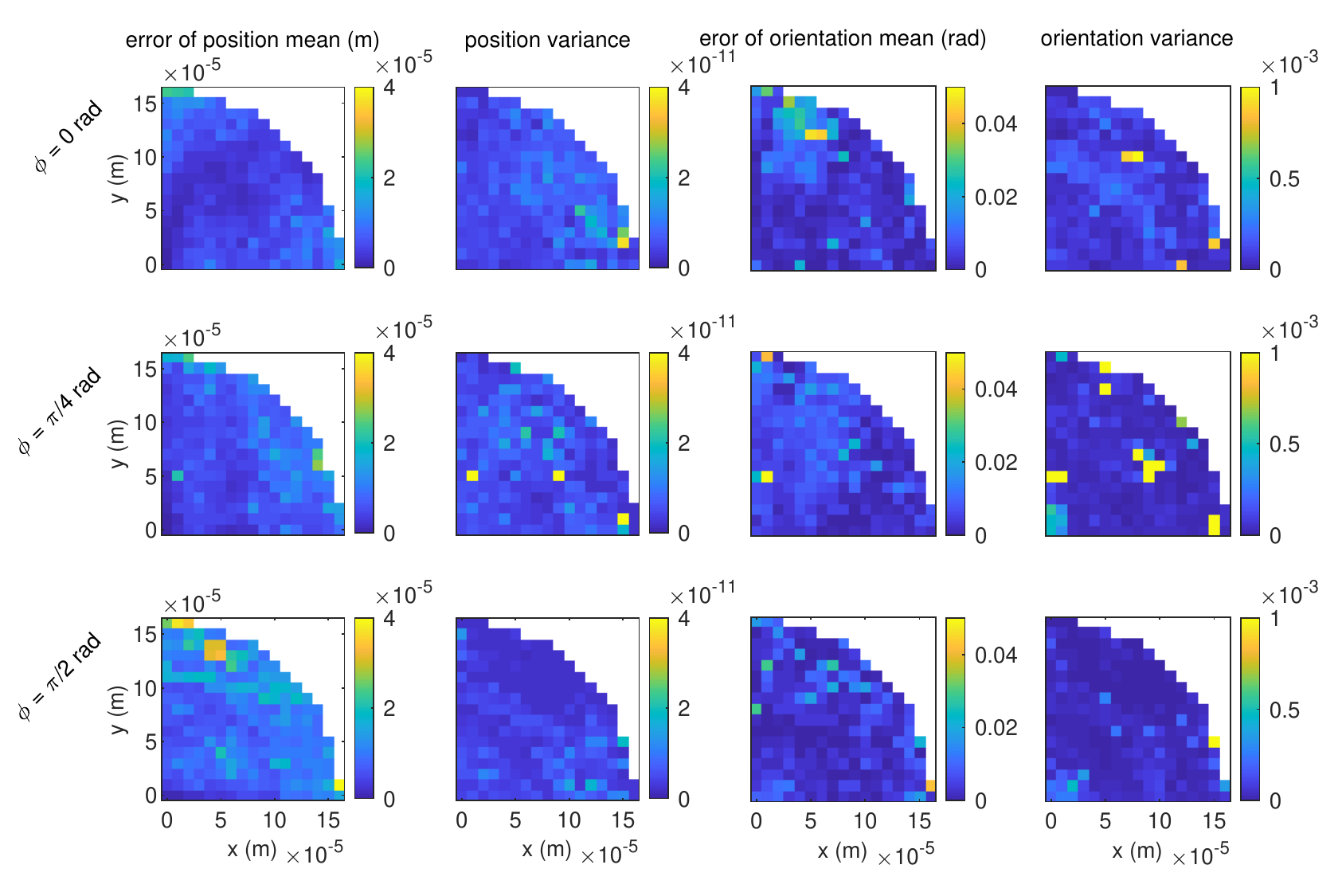}
  \caption{A set of measurements evaluating the precision (error of mean) and accuracy (variance) of the presented micro-manipulation. The object was repeatedly steered from an initial central position ($x=y=\SI{0}{\micro\meter}$) to a grid of positions inside a radius of approximately \SI{150}{\micro\meter} above one quarter of the manipulation area (the electrode array is symmetric). This was repeated for three different orientations of the object.}
  \label{fgr:precision_and_accuracy}
\end{figure*}

We evaluated the precision and accuracy of the manipulator by a set of experiments, in which we steered the ``S/Z''-shaped object to a grid of predefined positions and orientations over one quarter of the manipulation area (we utilize the symmetry of the electrode array to reduce the number of experiments to be done). Every experimental run started from a location $x=y=\SI{0}{\micro\meter}$, and $z=\SI{100}{\micro\meter}$ with an orientation $\psi=\theta=\phi=\SI{0}{\radian}$. An experiment ended either if the object achieved its goal position with a small enough tolerance (than it was kept there for additional \SI{3}{\second}) or a time limit of \SI{20}{\second} elapsed. The measured positions and orientation from the last \SI{3}{\second} of every experiment were then used to calculate the error of the mean value of position and orientation, which signify the accuracy of the manipulation and the variance of measured positions and orientations, which signify its precision. The results are shown in~\cref{fgr:precision_and_accuracy}. The accuracy of positioning was most of the time around \SI{5}{\micro\meter}, while the accuracy of orientation was around \SI{0.01}{\radian}. As can be noticed, the orientation of the object influences the accuracy of positioning. In the case of $\phi=\SI{0}{\radian}$, there is a notably smaller error along the $y$-axis, while for $\phi=\frac{\pi}{2}\,\mathrm{rad}$ an error along the $x$-axis was smaller. Generally, the closer the boundary of the object approaches the electrodes, the less accurate the manipulation is.

Regarding the manipulation speed, the object achieved its final position (with the tolerance noted above) in \SI{4.6}{\second} on average, which corresponds to a manipulation speed of $\sim\SI{22.1}{\micro\meter\per\second}$.

\section{Conclusions}

We described a non-contact micro-manipulation technique based on dielectrophoresis capable of arbitrary positioning and orienting micro-object of various shapes.

The object is automatically tracked in a video stream, and the controller using real-time optimization-based inversion of the gDEP model finds the most suitable phase-shifts of the voltage signals used for actuation.

Experiments performed on a quadrupolar electrode array demonstrated the micro-manipulation capabilities of the proposed approach. The analysis of manipulation accuracy showed that the mean error in position is around \SI{5}{\micro\meter}, while the error in orientation is around \SI{0.01}{\radian}. A video showing some of the experiments is available online at~\url{https://youtu.be/SBepX_Xk1BM}.

Our solution enables us to manipulate with objects of arbitrary shapes being made from a broad class of polarizable materials and being even inhomogeneous. With a higher number, different sizes, and arrangements of electrodes, the technique is readily extensible to simultaneous manipulation with multiple and even much smaller objects. Since the manipulation is non-contact, it could also be used in closed microfluidic systems for biology or micro-assembly applications. The latter is our current research direction.



\section*{Acknowledgment}
This work has been supported by the Czech Science Foundation within the project P206/12/G014 (Center for advanced bioanalytical technology). Thanks should also go to the French RENATECH network and its FEMTO-ST technological facility for manufacturing the used electrode array and micro-objects as part of our previous collaboration.


\ifCLASSOPTIONcaptionsoff
  \newpage
\fi



\bibliographystyle{IEEEtran}
\bibliography{IEEEabrv,rsc}

%



%

\begin{IEEEbiography}[{\includegraphics[width=1in,height=1.25in,clip,keepaspectratio]{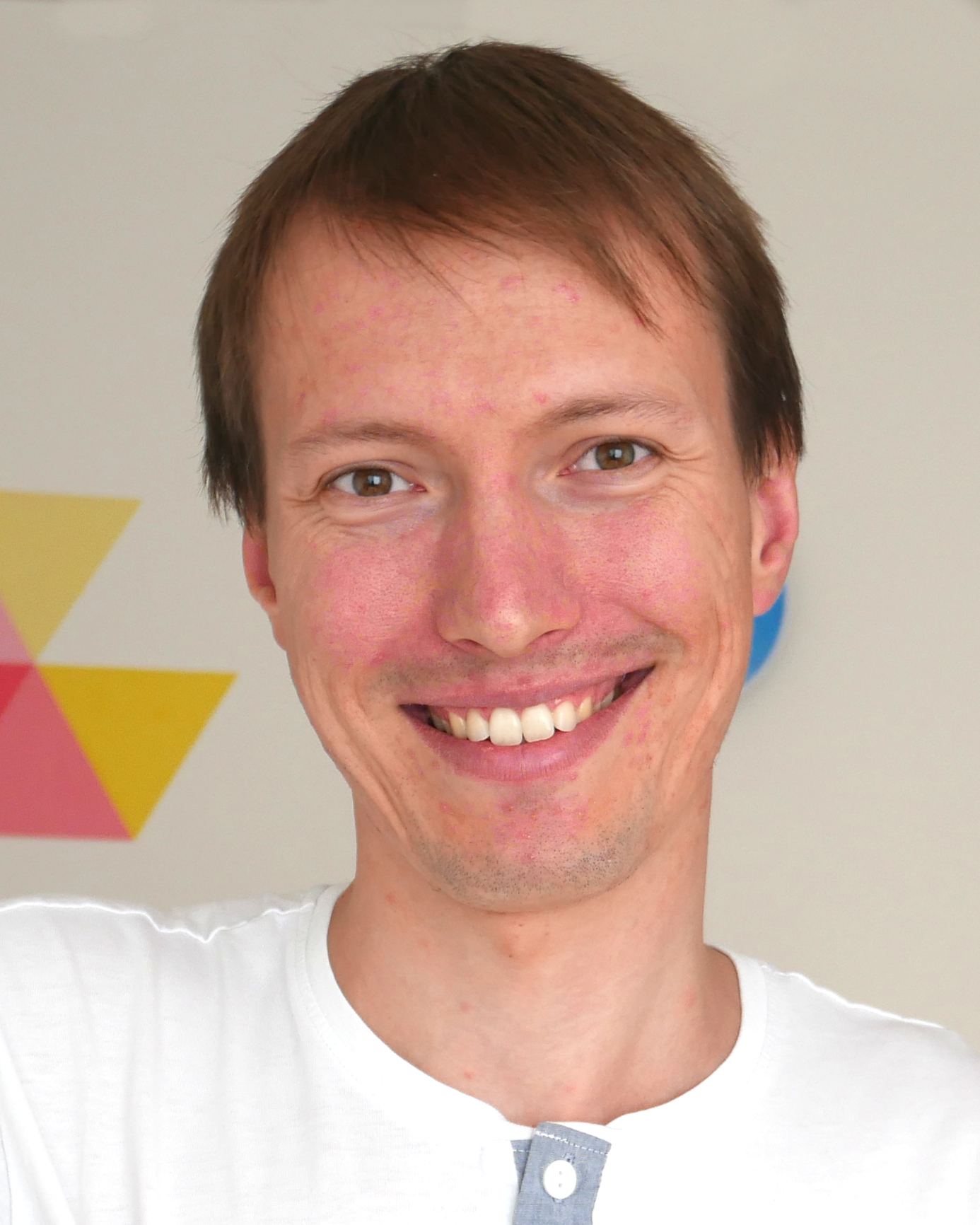}}]{Tom\'a\v{s}~Mich\'alek}
received his Ing. (~M.Sc.) degree in Cybernetics and Robotics from the Czech Technical University in Prague, Prague, Czech Republic, in 2015. He is currently working toward the Ph.D. degree in Control Engineering and Robotics.

His research interests include mathematical modeling and control design for noncontact micromanipulation via dielectrophoresis and electrorotation.
\end{IEEEbiography}


\begin{IEEEbiography}[{\includegraphics[width=1in,height=1.25in,clip,keepaspectratio]{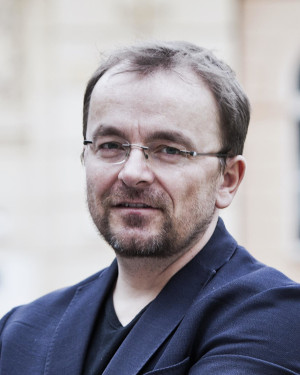}}]{Zden\v{e}k~Hur\'ak}
received the Ing. (~M.Sc.) degree in aerospace electrical engineering (summa cum laude) from Military Academy, Brno, Czech Republic, in 1997. He received the Ph.D. degree in cybernetics and robotics from the Czech Technical University, Prague, Czech Republic, in 2004. Currently he is an associate professor of control engineering at Faculty of Electrical Engineering, Czech Technical University in Prague, Czech Republic. His research focus is on computational methods for optimal, robust and distributed control and their applications in electromechanical systems, including non-contact (micro)manipulation using dielectrophoresis and magnetophoresis, more info on his group webpage \url{http://aa4cc.dce.fel.cvut.cz}.
\end{IEEEbiography}

\vfill




\end{document}